# Blockchain in the Eyes of Developers

He Jiang[1,2], Dong Liu[1], Zhilei Ren[1], Tao Zhang[3]
[1]*School of Software, Dalian University of Technology*
[2]*School of Computer Science & Technology, Beijing Institute of Technology*
[3]*College of Computer Science and Technology, Harbin Engineering University*

***ABSTRACT.** The popularity of blockchain technology continues to grow rapidly in both industrial and academic fields. Most studies of blockchain focus on the improvements of security, usability, or efficiency of blockchain protocols, or the applications of blockchain in finance, Internet of Things, or public services. However, few of them could reveal the concerns of front-line developers and the situations of blockchain in practice. In this article, we investigate how developers use and discuss blockchain with a case study of Stack Overflow posts. We find blockchain is a relatively new topic in Stack Overflow but it is rising to popularity. We detect 13 types of questions that developers post in Stack Overflow and identify 45 blockchain relevant entities (e.g., frameworks, libraries, or tools) for building blockchain applications. These findings may help blockchain project communities to know where to improve and help novices to know where to start.*

## Introduction

The public interest of blockchain has been increasing since it was proposed[1]. Blockchain is essentially a public ledger of all committed transactions or digital information shared by all participants. One of the most popular examples of blockchain applications is Bitcoin, a cryptocurrency by which transactions could happen without any third party[2]. Nevertheless, the ability of blockchain is far beyond Bitcoin. It enables the refinement of existing techniques and the appearance of new applications[3].

From the technical point of view, blockchain is a technique involving several fields of computer science, e.g., networking, cryptography, and information security. Many studies of blockchain have been conducted in these fields, for example, protocols design, self-certification, and vulnerability protection[1]. However, few of them focus on problems from front-line developers' perspective. *What is the state of blockchain technology in software development communities? Is it also a hot topic? What kinds of problems do developers encounter about blockchain? What programming frameworks, tools, libraries, or techniques can be used in blockchain relevant tasks?* These questions are still open to us. By answering the questions, novice developers could know what resources to exploit to support their blockchain relevant programming tasks and blockchain project communities could better acknowledge how to make blockchain technology more practical to front-line developers.

In this article, we conduct a case study by Stack Overflow, a popular technical Question and Answer (Q & A) website attracting millions of developers. We extracted all the blockchain relevant question posts and investigate how blockchain is discussed by developers there.

## Overview of the Study

In this study, we attempt to draw a profile of blockchain in the eyes of developers with Stack Overflow as a point of penetration. In brief, the motivation of this study contains three parts. First, we want to evaluate the popularity of blockchain, or to check whether blockchain is hot. To this end, we use the frequency of blockchain relevant questions in Stack Overflow as an index. Second, we want to acknowledge what are the frequently encountered problems about blockchain. Hence, we analyze the blockchain relevant questions in Stack Overflow to find the common terms. Third, we also want to know what frameworks, tools, libraries, or techniques can be used by developers in blockchain relevant tasks. So we try to identify these entities from tags of Stack Overflow posts. In summary, the study is to answer the following three questions:

*Q1: How frequently blockchain relevant questions are posted in Stack Overflow?*

*Q2: What are the common problems when blockchain is involving?*

*Q3: What blockchain specialized entities (programming relevant platforms, frameworks, tools, libraries, techniques, etc) do there exist?*

With these questions, the study is conducted according to the following steps. At first, we extract the question posts in which the string "blockchain" is mentioned. An evolvement analysis of these posts is conducted. Next, we manually categorize these extracted question posts into different types in perspective of developers' demands. The description of each type is summarized further. Then, blockchain relevant tags of the extracted question posts are selected. After that, these tags are identified as commonly used frameworks, tools, libraries, or techniques related to blockchain. Figure 1 outlines the process of this study.

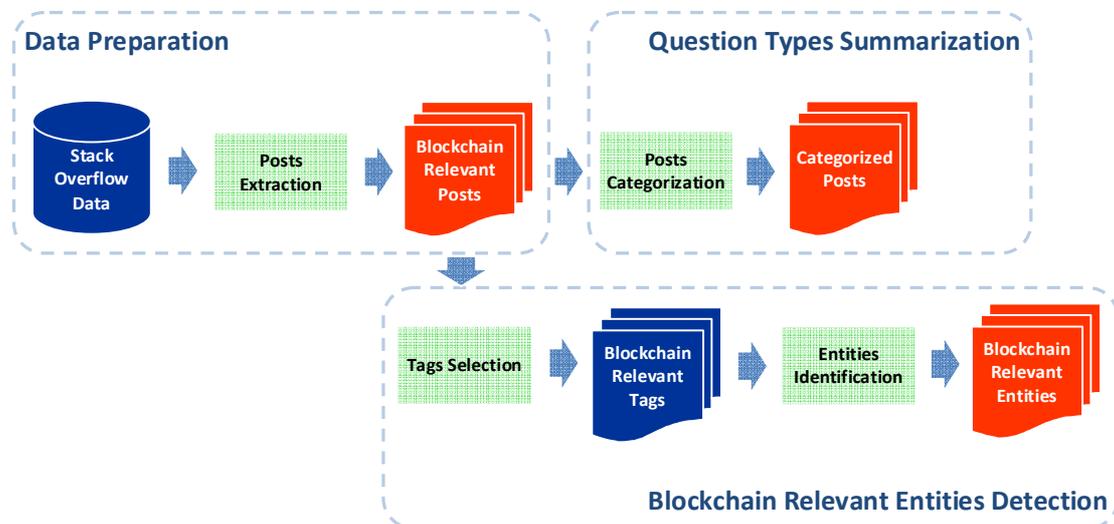

Figure 1. The process of this study. Blockchain relevant entities indicate concepts specialized in Blockchain technology such as platforms, frameworks, or techniques.

## Data Preparation

**Posts Extraction**

The Stack Overflow data is publicly available on Internet Archive (https://archive.org/details/stackexchange). In this study, we download the file *Posts.xml* which contains all the posts of Stack Overflow. The file contains 15,483,377 question posts which were posted from July 2008 to March 2018. We parse the file to obtain post metadata, for example, the post title, the post body, the tags, and the score.

With these data, we extract the question posts in which "blockchain" is mentioned. Specifically, we first remove all the code snippets and hyperlink text (enclosed in <a href="..."> HTML tags) from the bodies of all the question posts. Then, we go through the titles, bodies, and tags of all the question posts. A question post is extracted as blockchain relevant if the string "blockchain" or "block[^a-z]chain" appears in at least one of these three fields, where "[^a-z]" represents a character but not a letter. Using these heuristics, 1,785 question posts are identified as blockchain relevant.

**Evolvement of the Blockchain Relevant Posts**

Among all 15,483,377 question posts in Stack Overflow, only 1,785 of them are blockchain relevant. We suppose that blockchain is an emerging concept and going to rise. To prove this idea, we investigate the evolvement of the blockchain relevant question posts. Figure 2 presents the number of blockchain relevant question posts per month. The Stack Overflow website was operative as early as July 2008, but blockchain relevant question posts were not mentioned for a long time at the beginning. The first blockchain relevant question was posted in January 2012. The time span before that date is omitted in the figure. The data of March 2018 is eliminated as it is not complete.

We have two observations from the evolvement. On the one hand, blockchain is an ongoing topic in Stack Overflow programming practice. The monthly number of blockchain relevant question posts shows a clearly increasing trend especially since January 2016. Moreover, the increasing speed tends to increase, too. That means more and more programming issues about blockchain are raised and discussed. Although the current number of blockchain relevant question posts is relatively small, we believe that it could rise fast in the future. On the other hand, the application of blockchain is lagging behind the theory of blockchain. The idea of blockchain was conceptualized in 2008[4]. But only still 2012, it emerged in Stack Overflow, where blockchain is viewed in the application perspective. There is a time delay between theory and application. However, blockchain has attracted the attentions from programming communities and blockchain relevant questions are continuously growing.

**A1. Currently, blockchain relevant questions only occupy a small part of all Stack Overflow posts, but the monthly number is increasing over time (Figure 2).**

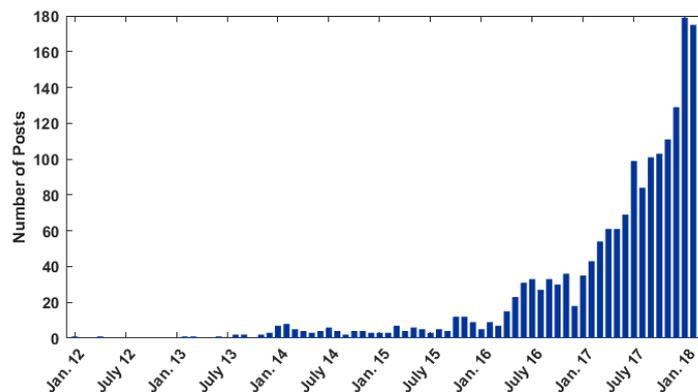

Figure 2. Evolvement of blockchain relevant question posts.

## Question Types Summarization

**Posts Categorization**

To give a summary of what tasks developers face with, we categorize the blockchain relevant question posts into different types. As the extracted question post set is not very large, and for the sake of accuracy, we do this work manually.

To make the results more understandable, we do not focus on technical details but the demands of askers. A question post is labeled with a unique type which reflects the main purpose of the asker. For example, in a post, the asker writes "*I want to query the blockchain to find the moment when an unconfirmed transaction comes through for a particular Bitcoin address*" (#25882458). That is, the asker wants to access the blockchain to obtain some information of transactions. Then we may label the post as "access". In another example, "*what's the difference between distributed hashtable technology and the Bitcoin blockchain?*" (#26415908), the asker just wants to discriminate the two concepts but does not leak the practical requirements. Then the post could be labeled as "discussion".

With this consideration, we categorize the posts following the iterative process of mobile app reviews tagging[5]. Specifically, for each of the blockchain relevant question post, we label it with an existing type. If a question post does not match any existing types, then we add a new type to the existing types and relabel the blockchain relevant question post from the beginning. Finally 13 types of blockchain relevant question posts are detected.

**Types of the Blockchain Relevant Question Posts**

For the 13 types of blockchain relevant question posts, we summarize their substances and give a brief description and an example question post for each type. The details are shown in Figure 3.

As a question post is labeled with a unique type, the blockchain relevant question posts are grouped into 13 clusters. If a large number of question posts belong to the same type, then this type of issues are frequently encountered and should be paid more attention to. The number of question posts a type involves (cluster size) reveals the popularity of this type. To find the popular types, we rank the 13 types according to the numbers of question posts and the results are displayed in Figure 4 (a).

The top three types are *Configuration*, *Deployment*, and *Discussion*, which account for more than 40 percent of all the posts. Questions of *Configuration* are about how to configure the environment to work well. Many askers reported to get errors during the configuring process, such as post#38572046: *Error Running custom blockchain (hyperledger) application with security enabled*. Similarly, *Deployment* is about how to set up or install the blockchain system, such as post#47194702: *How can i integrate etherium or waves platform wallet on my site?* Among all the askers of these two types of questions, a number of them are in the learning phase and do not have certain tasks. Askers of 113 question posts declared that they tried to follow the steps provided in the tutorials. Questions of *Discussion* are related to more general topics. The askers may want to talk about the ideas (e.g., *are there already other practical uses for blockchain?* #23846848), understand the mechanisms (e.g., *Where are chaincodes executed?* #36910077), resolve their puzzles (e.g., *bitcoins, how can someone receive much less than have spent?* #37772139), or verify their thoughts (e.g., *Is blockchain a decentralised* *database?* #38558703). These facts imply that the application of blockchain is still in the initial phase. Many developers are confused with the principles of blockchain and many developers have set out to work with blockchain.

| Type | Description |
|---|---|
| Access | Retrieve or query information from blocks, cryptocurrency accounts, addresses, etc. |
| | Example post: *How to retrieve a value from a Blockchain using PHP? (#21331341)* |
| Application | Design decentralized applications by leveraging blockchain technology |
| | Example post: *How can blockchains be used in audit trails? (#45593747)* |
| Authentication | Authenticate the participants by verifying the certifications or signatures |
| | Example post: *Using the Blockchain.info API to verify bitcoin signatures in C# (#34364767)* |
| Configuration | Configure blockchain systems to make them work correctly |
| | Example post: *Error Running custom blockchain (hyperledger) application with security enabled (#38572046)* |
| Connection | Connect to blockchain systems or peer networks |
| | Example post: *How to correctly respond to remote server? (#21520358)* |
| Contract | Design or implement the logic of smart contracts |
| | Example post: *How to create contracts on Ethereum block-chain in python? (#38975431)* |
| Cryptography | Achieve encryption or decryption of payloads, or calculate hash values |
| | Example post: *How to decrypt the transaction payload when confidentiality is turned on? (#38143344)* |
| Deployment | Deploy or install blockchain relevant frameworks on their environments |
| | Example post: *How can I integrate etherium or waves platform wallet on my site? (#47194702)* |
| Discussion | Post the opinions, discuss the puzzle about blockchain technology |
| | Example post: *What's the difference between distributed hashtable technology and the bitcoin blockchain? (#26415908)* |
| Processing | Parse, transfer, or store the blockchain data |
| | Example post: *Bitcoin: parsing Blockchain API JSON in PyQT (#26479696)* |
| Programming | Clarify programming details of blockchain relevant technology |
| | Example post: *Programming on Solidity setName function (BitDegree) (#48362797)* |
| Protocol | Implement the protocols between peers, i.e., consensus, fault tolerance |
| | Example post: *How exactly to make consensus when Chaincode has coding block of authority or event? (#38215761)* |
| Transaction | Make transactions, receive or send cryptocurrency between accounts |
| | Example post: *Simple 1-to-1 Bitcoin Transaction with Bitcore (#46528602)* |

Figure 3. The 13 types of blockchain relevant question posts.

Then we evaluate the 13 types in another perspective. A question post can be evaluated with different metrics. For example, *score* measures the quality of the post, *viewcount* measures the number of visitors, and *favoritecount* measures how many developers mark it as favorite. A post may get high values of these metrics under the following scenarios:

the question is just what many other developers want to ask; the topic of the question is what many other developers concern; the question is related to the jobs what many other developers are doing. In brief, many other developers are interested in the question. To find the types developers interested in, we rank the 13 types according to the average score, viewcount, and favoritecount of posts they involve. The results are shown in Figure 4 (b) – (d).

We note that the three metrics of question posts belonging to *Contract* and *Protocol* are all in top ranks. Questions of *Contract* are about how to develop or implement the smart contract, that is, to contribute, verify or implement the negotiation or performance of the contract[6]. In brief, they are about the business logics of blockchain, e.g., post#42768684: *How to Destroy Tokens/Coins in a Smart Contract?* Questions of *Protocol* are about how to deal with protocols to make blockchain work correctly. That is, they are about the underlying logics of blockchain, e.g., post#38215761: *How exactly to make consensus when Chaincode has coding block of authority or event?* In general, the types that developers are interested in are all related to the operating mechanisms of blockchain.

There are some implications of the findings. The official organizations of existing blockchain platforms, frameworks, or libraries may provide more manuals, tutorials, or books to facilitate the novice developers to handle these techniques. The practitioners of blockchain may write more informal reading materials to make the masses understand the fundamental principles of blockchain more easily. Moreover, the blockchain communities may set up specialized forums or websites so that the blockchain enthusiasts could discuss or exchange their ideas more conveniently.

**A2. Questions about blockchain can be roughly categorized into 13 types (Figure 3). Among them, questions about configuration, deployment, discussion, and mechanisms (contract or protocol) of blockchain are notable.**

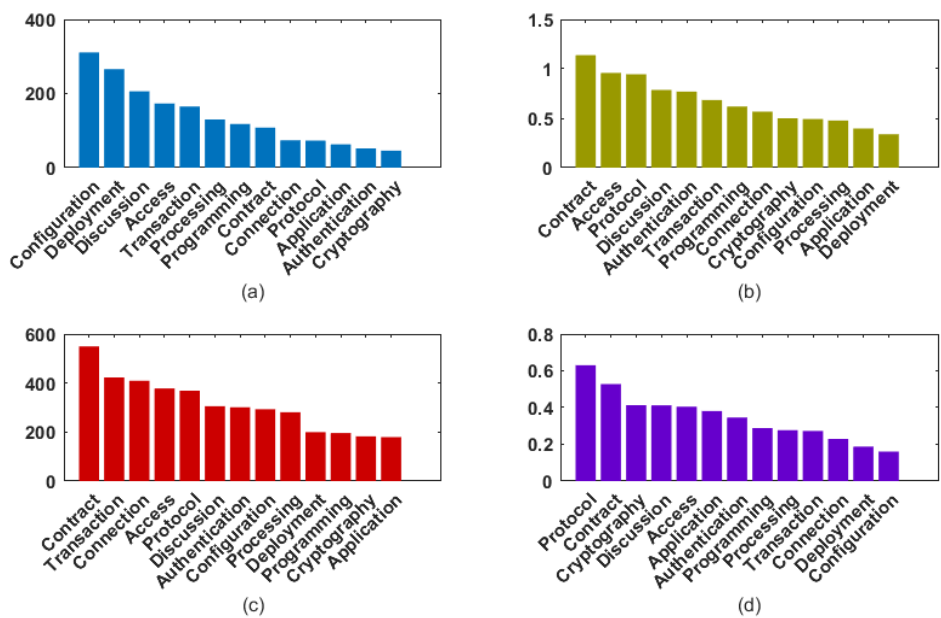

Figure 4. The ranks of 13 types according to different metrics.

# Blockchain Relevant Entities Detection

**Tags Selection**

In Stack Overflow, a question post contains up to five tags. The tags can usually reveal the main technologies the question post involves[7], and they can usually be identified as named entities, e.g., operating systems, APIs, and platforms[8]. Therefore, we attempt to find commonly used frameworks, tools, libraries, or techniques which are relevant to blockchain from the tags of the blockchain relevant question posts.

To select blockchain relevant tags, we leverage association rule mining. Generally, if a tag frequently appears and only appears in the blockchain relevant question posts, then this tag is highly possible to be specialized in blockchain. There are usually two measures of each tag: *support* indicates the frequency the tag appears in the blockchain relevant question posts; *confidence* represents the proportion of the appearances of the tag in the blockchain relevant question posts to the total appearances of this tag.

The common process of association rule mining is to set thresholds for the two measures respectively and select the tags whose measures are above the thresholds. However, blockchain is a relatively new topic in Stack Overflow and blockchain relevant tags may appear very few times, for instance, only one time. Therefore, we set the threshold of *support* to be zero, i.e., all the tags which appear in the blockchain relevant question posts are considered, so that more potential blockchain relevant tags can be selected.

For *confidence*, we sort the values of this measure of all the tags in a descending order and draw a curve of the sorted values with their ranks. Then we find the *knee* of the curve[9]. That is, in the curve, the value decreases steeply before the *knee* and get smooth after the *knee*. The value at the *knee* is chosen as the threshold of *confidence*. In this way, 49 tags are selected as their measures exceed the corresponding thresholds.

**Entities Identification**

For each of the blockchain relevant tag, we check the tag's descriptions in Stack Overflow to determine what the entity the tag represents. For example, in Stack Overflow, the tag "web3js" is described as "*The web3.js library is a collection of modules which contain specific functionality for the ethereum ecosystem allowing to ...*" It implies that the tag "web3js" represents web3.js, which is a JavaScript library used for Ethereum.

Nevertheless, descriptions of some tags can not be found in Stack Overflow. Under this condition, we search the tag on the Internet and check which descriptions are related to blockchain. Fortunately, there are no ambiguities. That means, a unique entity can be deduced from the information of the tag which is related to blockchain.

When the entities of all the blockchain relevant tags are identified, the blockchain relevant frameworks, tools, libraries, or techniques are clear to us.

**Blockchain Relevant Entities**

The identified entities of the 49 selected tags are analysed and summaried. Sometimes two tags represent the same entity, i.e., the two tags duplicate each other. Then we merge them into one. There are 45 unique entities in total. Figure 5 draws the *word cloud* (https://wordart.com/create) of the 45 entities. The size of an entity name indicates the number of blockchain relevant question posts containing this entity as a tag. There is undeniably that the biggest one is *Blockchain*. Moreover, we give each entity a summary to show its connotations. If the readers are interested in any of these entities, then more information could be got by retrieving on the Internet with their full names as keywords.

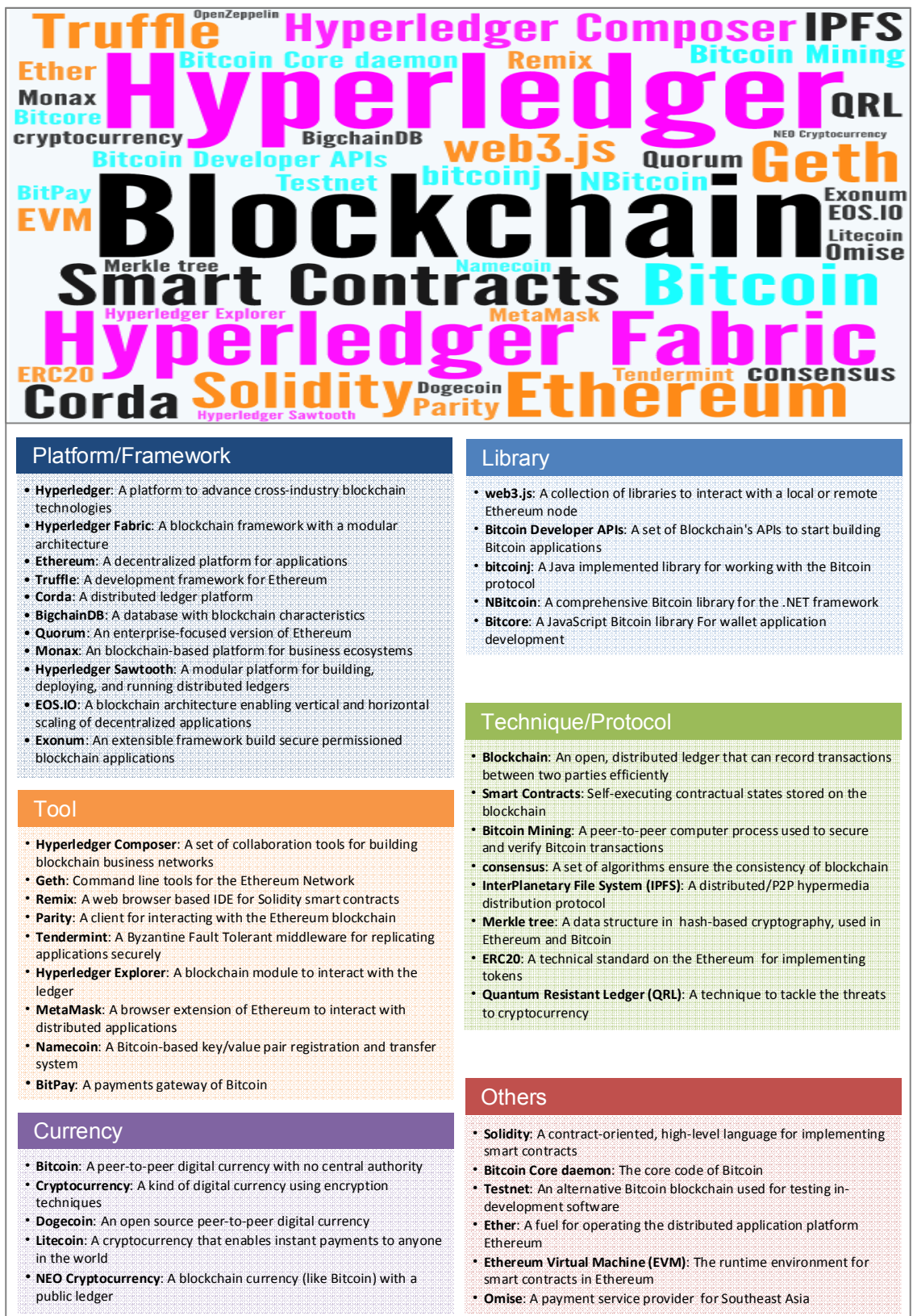

Figure 5. The word cloud of identified entities relevant to blackchain.

These entities cover a wide range of categories. A large part of them belong to *Platform/Framework* which helps users to build their blockchain systems in different aspects. For example, *Ethereum* allows users to build and run decentralized applications operated through blockchain techniques; *Corda* allows businesses to transact directly in strict privacy; and *BigchainDB* allows developers and enterprise to deploy blockchain applications with a blockchain database. Entities in *Tool* are softwares or systems for certain uses, e.g., *Parity*, *Hyperledger Explorer*, and *MetaMask* are for interaction with blockchains. *Library* is a set of APIs for developers to build applications associated with blockchain or Bitcoin. We note that all the listed *Platforms/Frameworks*, *Tools*, and *Libraries* are freely available. Among all the categories, *currency* means cryptocurrency, a specialized one for blockchain. A cryptocurrency is a digital currency in which cryptography techniques are used to regulate its generation and transfer[10]. *Bitcoin* is the first decentralized cryptocurrency and we find cryptocurrencies other than *Bitcoin*, for example, *Dogecoin* and *Litecoin*. Technique/Protocol is referred to as technique, protocol, standard, or concept used in blockchain, or just blockchain itself. Moreover, there are other categories, e.g., *Solidity*, an advanced programming language.

For all 45 entities, we find three big groups: *Hyperledger*, *Ethereum*, and *Bitcoin*. In the word cloud (Figure 5), the font colors of the entity names which are in the three groups are set to be magenta, orange, and cyan, respectively. It means that these entities are relevant to *Hyperledger*, *Ethereum*, or *Bitcoin*. The numbers of involving posts of the three groups are 638, 432, and 320, which are relatively large numbers compared with the total number of blockchain relevant question posts (i.e., 1,785). Therefore, we can say that *Hyperledger* and *Ethereum* are two most concerned about blockchain platforms in Stack Overflow. Moreover, unsurprisingly, *Bitcoin* is discussed frequently.

With the identified blockchain relevant entities, the beginners could have a glimpse of blockchain relevant concepts. Especially, they could use these open source frameworks, tools or libraries to build their blockchain systems, for example, they could begin with the most popular platforms in Stack Overflow, *Hyperledger* or *Ethereum*.

**A3. We identify 45 blockchain relevant entities (Figure 5) in Stack Overflow. The entities in three groups, which are relevant to Hyperledger, Ethereum, and Bitcoin, are frequently mentioned.**

## Conclusion

Blockchain is going to be a hotspot in both industrial and academic fields. Our study investigates what are mostly concerned in the perspective of software developers. We find that most developers in Stack Overflow are just in stage of learning and attempt. Many of them want to know how to correctly deploy and configure the blockchain systems with existing platforms or frameworks and they are also interested in the mechanisms of blockchain. In general, the three primary topics discussed in Stack Overflow are *Hyperledger*, *Ethereum*, and *Bitcoin*.

With these findings, the blockchain project communities could acknowledge the directions for making more practical blockchain techniques and novices could know what resources to exploit to build their first blockchain projects. In the future, we plan to expand our study to other Q & A platforms, forums, or blogs.